\documentclass[10pt]{article}

\usepackage[backend=bibtex,style=nature]{biblatex}
\usepackage{standalone} 
\usepackage[colorlinks,bookmarks=false]{hyperref} 
\usepackage{enumerate} 
\usepackage{enumitem} 
\usepackage{subcaption} 
\usepackage{amsfonts} 
\usepackage{braket} 
\usepackage{amssymb} 
\usepackage{amsmath} 
\usepackage{graphicx} 
\usepackage{caption} 
\usepackage{authblk} 
\usepackage{framed} 
\usepackage{multicol} 

\usepackage[
  margin=1in,
  includefoot,
  footskip=30pt,
]{geometry}
\usepackage{layout}

\usepackage[table,xcdraw,svgnames]{xcolor}
\AtBeginDocument{%
  \hypersetup{
    citecolor=NavyBlue,
    linkcolor=Red,   
    urlcolor=Purple}}

\bibliography{LLP_library}

\newbibmacro{string+doi}[1]{%
  	\iffieldundef{doi}{
  		\iffieldundef{url}{#1}{\href{\thefield{url}}{#1}}
	}{\href{http://dx.doi.org/\thefield{doi}}{#1}}}
\DeclareFieldFormat{title}{\usebibmacro{string+doi}{\mkbibemph{#1}}}
\DeclareFieldFormat[article]{title}{\usebibmacro{string+doi}{\mkbibquote{#1}}}

\newcommand{\ispath}[2]{\langle #1 \leftrightarrow #2\rangle}
\newcommand{\emphrm}[1]{\mathrm{\emph{#1}}}
\newcommand{\threed}[3]{#1 \!\times\! #2 \!\times\! #3}
\newcommand{\twod}[2]{#1 \,\!\!\times\!\!\, #2}

\begin{document}
\title{Physical-depth architectural requirements for generating universal photonic cluster states.}
\date{}
\author[1,2]{Sam Morley-Short\thanks{\href{sam.morley-short@bristol.ac.uk}{sam.morley-short@bristol.ac.uk}}}
\author[3]{Sara Bartolucci}
\author[1,3,4]{Mercedes Gimeno-Segovia}
\author[3]{Pete Shadbolt}
\author[1]{Hugo Cable\thanks{\href{hugo.cable@bristol.ac.uk}{hugo.cable@bristol.ac.uk}}}
\author[3]{Terry Rudolph}

\renewcommand\Authfont{\fontsize{12}{14.4}\selectfont}
\renewcommand\Affilfont{\fontsize{9}{10.8}\itshape}

\affil[1]{Quantum Engineering Technology Labs, H.\ H.\ Wills Physics Laboratory and Department of Electrical and Electronic Engineering, University of Bristol, BS8 1FD, UK}
\affil[2]{Quantum Engineering Centre for Doctoral Training, H.\ H.\ Wills Physics Laboratory and Department of Electrical and Electronic Engineering, University of Bristol, Tyndall Avenue, BS8 1FD, UK}
\affil[3]{Department of Physics, Imperial College London, London SW7 2AZ, UK}
\affil[4]{Institute for Quantum Science and Technology, University of Calgary, Alberta T2N 1N4, Canada}

\maketitle
\vspace{-30pt}
\begin{abstract}
	Most leading proposals for linear-optical quantum computing (LOQC) use cluster states, which act as a universal resource for measurement-based (one-way) quantum computation (MBQC).
	In \emph{ballistic} approaches to LOQC, cluster states are generated passively from small entangled resource states using so-called fusion operations.
	Results from percolation theory have previously been used to argue that universal cluster states can be generated in the ballistic approach using schemes which exceed the critical threshold for percolation, but these results consider cluster states with unbounded size.
	Here we consider how successful percolation can be maintained using a physical architecture with fixed physical depth, assuming that the cluster state is continuously generated and measured, and therefore that only a finite portion of it is visible at any one point in time.
	We show that universal LOQC can be implemented using a constant-size device with modest physical depth, and that percolation can be exploited using simple pathfinding strategies without the need for high-complexity algorithms.
\end{abstract}

\section{Introduction}

Within the last decade, great progress has been made in the theoretical field of quantum computer architectures.
Modern fault-tolerant schemes rely on the use of many error-prone \emph{physical} qubits to create individual \emph{logical} qubits with fewer errors.
Whilst we understand these methods of abstraction theoretically, implementing them in reality is not a trivial task when experimental constraints are applied.
The study of quantum computation architectures must therefore incorporate both an understanding of high-level theoretical models and experimental limitations.

While there are many attractive aspects of photonic qubits, utilising them for linear-optical quantum computation (LOQC) presents some unique architectural challenges \cite{Rudolph2016}.
Most significantly, LOQC suffers from a lack of deterministic entangling gates, with initial proposals requiring large resource overheads to compensate \cite{Knill2001,Kok2007}.
However, the main challenge for modern LOQC architectures \cite{Nielsen2004,Browne2005,Varnava2008,Campbell2009,Kieling2007,Zaidi2015,Gimeno-Segovia2015,Li2015,Rohde2015,Uskov2015,Gimeno-Segovia2016} remains the generation and utilisation of highly-entangled resource states.
This is now generally addressed within the paradigm of cluster states \cite{Raussendorf2001,Raussendorf2003} applied to LOQC \cite{Nielsen2004} and the use of entangling fusion gates \cite{Browne2005,Grice2011,Ewert2014}.

One particularly appealing approach to LOQC uses ideas from percolation theory as first proposed in \cite{Kieling2007}.
The main idea is to passively entangle small resource states (also called microclusters), using fusion gates, to generate a large cluster state which can enable universal quantum computing.
The cluster state which is generated corresponds to a random graph on a geometric lattice with missing sites and bonds.
By using schemes which exceed the critical threshold for percolation on the lattice \cite{Kieling2007,Zaidi2015,Gimeno-Segovia2015,Pant2017}, a cluster state which supports universal quantum computation can be guaranteed.
A lattice of logical qubits can then be identified using methods such as renormalisation as given in Ref.\ \cite{Kieling2007}, or the lattice concentration algorithm of Ref.\ \cite{Browne2008}.
The main virtue of using the percolation approach to LOQC is that it enables \emph{ballistic} architectures that sidestep requirements for extensive adaptive switching networks, which are technologically very challenging \cite{Silverstone2016}.

In this work, we address a vital question that must be addressed for any high-level LOQC architecture based on percolation: Can successful percolation be sustained using a physical device of fixed finite size, and what size (cross-section and depth) of percolating cluster state must be kept \emph{online} at any point in time to do so?
The methods we use to answer this question differ from conventional treatments of percolation, and are based on pathfinding algorithms which must exploit information in real-time about the outcomes of recent fusion operations.
We assume that photons making up the percolating cluster state can only be kept online for modest periods using optical delays, which provide limited \emph{lookahead} capability before measurements must be performed on the photons.
Our analysis can have implications for all aspects of LOQC architecture by impacting hardware specifications at the component level.
Specifically, this work presents three key results: i) spanning paths can exist on extremely elongated blocks of edge-percolated cluster state lattice, but only when the cross-sectional side length exceeds some minimum length set by the lattice edge probability; ii) an LOQC device with a physical-depth of only 10-20 layers is sufficient to produce MBQC qubit channels (within a loss- and error-less LOQC architecture model); iii) long-range limited-lookahead pathfinding can be achieved with algorithms with minimal complexity, thereby reducing associated classical co-processing requirements for LOQC.

The structure of this work is as follows: In Sec.\ \ref{sec:LOQC} we briefly review recent work on percolation-based architectures for LOQC.
In Sec.\ \ref{sec:long-range_perc} we consider the minimum resource requirements of percolated cluster state lattices for producing long-range single-qubit channels.
In Sec.\ \ref{sec:LLAPF} we present the main results of our work, where we define the Random-node pathfinding process, conjecture a condition of pathfinding success and present results from numerical pathfinding simulations.
Sec.\ \ref{sec:implications} considers implications of the results presented for LOQC architectures, identifying key architectural trade-offs and specifications.
Finally, a selection of open questions for future work are presented in Sec.\ \ref{sec:open_questions}.

\section{Percolation-based architectures for LOQC} \label{sec:LOQC}

The fundamental challenge of LOQC is the construction of large graph states.
Graph states are a subset of stabilizer states \cite{Gottesman1997} that can be uniquely described by simple graphs (for a review of graph states see \cite{Hein2004}).
In this formalism, a graph $G(V,E)$ containing \emph{vertices} (or \emph{nodes}) $V$ and \emph{edges} (or \emph{bonds}) $E$, uniquely represents the state
\begin{align}
	\ket{\Psi_G} = \prod_{\langle i, j\rangle \;\in\; E} \!\!\!\!\!\emphrm{CZ}_{i,j}\bigotimes_{v \;\in\; V} \ket{+}_v ,
\end{align}
\textcolor{black}{where $\emphrm{CZ}_{i,j} = \ket{00}\bra{00}_{i,j} + \ket{01}\bra{01}_{i,j} + \ket{10}\bra{10}_{i,j} - \ket{11}\bra{11}_{i,j}$ and $\ket{+} = \frac{1}{\sqrt{2}}(\ket{0} + \ket{1})$.}
We specifically refer to graph states represented by regular lattices as \emph{cluster states}.
In LOQC, cluster states can be probabilistically built using two types of fusion gate \cite{Browne2005}.
Known as type-I and -II fusion gates, these gates destructively consume 1 and 2 photonic qubits respectively and on success produce entanglement between the remaining qubits in the clusters (and on failure the input qubits are subjected to single-qubit measurements).
Whilst type-I fusion consumes fewer qubits, it cannot herald photon loss, whereas type-II can herald such loss, but at the cost of consuming an extra qubit.
In standard operation, both gates operate with a 50\% success rate.
However, Type-II fusion can be \emph{boosted} to increase the success rate above 50\% through the consumption of additional auxiliary resources \cite{Ewert2014,Grice2011}.
For example, a success rate of 75\% can be achieved through either the consumption of a Bell pair or 4 single photons.

To overcome nondeterministic entangling gates, renormalization is used to produce an idealised lattice $\mathcal{L^*}$ from a coarse graining of some percolated lattice $\mathcal{L}$.
For example, in one common strategy, microcluster states are placed on the sites of a lattice and fusion gates of success probability $p_f$ are applied to produce entanglement between the centre qubits of adjacent microclusters.
Once $\mathcal{L}$ is constructed, a single \emph{central} qubit is identified on each renormalization block that is path-connected to central qubits of adjacent blocks by sets of \emph{path} qubits\footnote{For the renormalization of 2D lattices, a different method based on the identification of topological minors is also know \cite{Browne2008}, however this has yet to be extended to higher dimensional lattices.}.
As in MBQC protocols \cite{Raussendorf2001,Raussendorf2003}, all other qubits in the lattice are then removed by adaptive single-qubit measurements, thereby producing $\mathcal{L}^*$.
An example of this is depicted in Fig.\ \ref{fig:windows}, where a single-qubit MBQC channel is produced from the renormalization of a 2D lattice.

The size of blocks on $\mathcal{L}$ required for renormalization to a fixed $\mathcal{L^*}$ depends only on the percolation threshold $p_c$ of $\mathcal{L}$, as produced by the lattice's structure.
Reducing the overall resource requirements for a LOQC device therefore relies on producing a lattice with low $p_c$ without the need for high-degree and therefore costly microcluster resource states.
Initial work on renormalization identified cubic, diamond and pyrochlore lattices as potential candidates, requiring 7-, 5- and 4-qubit microcluster resources respectively \cite{Kieling2007}.
By extending a percolation approach to the generation of resource states, it was shown that both microcluster creation and fusion could be achieved from boosted fusion \cite{Grice2011,Ewert2014} of 3-photon GHZ states to produce a `brickwork' diamond lattice with $p_c < p_f$ \cite{Gimeno-Segovia2015} and pyrochlore \cite{Zaidi2015}.
Recently, this scheme was further generalised for higher-dimensional lattices and $n$-qubit microclusters \cite{Pant2017}.
After $\mathcal{L}$ has been constructed, renormalization can be abstracted to the graph-theoretical problem of finding crossing clusters on percolated lattices, which can be solved efficiently \cite{Hoshen1976}.

\begin{figure}[t]
	\centering
	\vspace{0pt}
	\includegraphics[width=\textwidth]{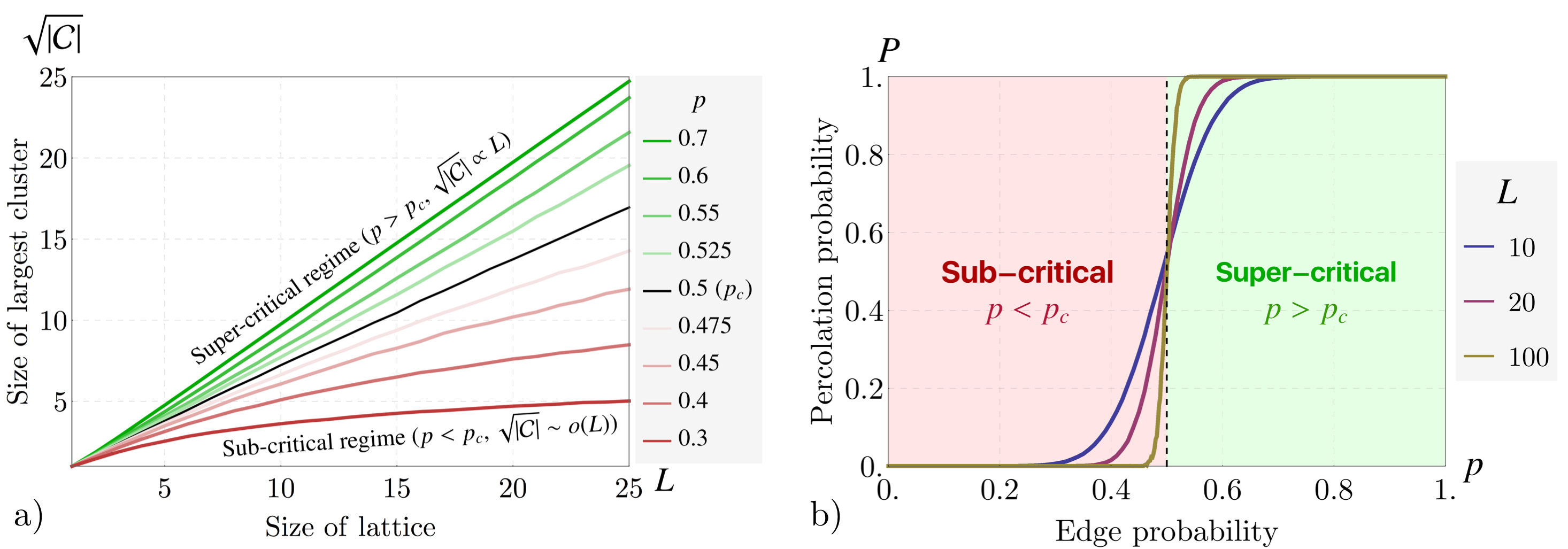}
	\caption{Percolation phenomena in the sub- and super-critical regime for $p<p_c$ and $p>p_c$ respectively, here produced by simulation of percolation on $\twod{L}{L}$ square lattices with edge probability $p$. a) The size of the largest connected component $\sqrt{|\mathcal{C}|}$ as a function of percolated lattice size $L$ (where $|\mathcal{C}|$ is the number of nodes in the connected component $\mathcal{C}$). For sub-critical percolation when $p<p_c$, the size of the largest connected component scales sub-linearly $\sqrt{|\mathcal{C}|} \sim o(L)$, whereas for super-critical percolation when $p>p_c$, the size of the largest component is proportional to the lattice size $\sqrt{|\mathcal{C}|} \propto L$. b) The probability of percolation $P$ as a function of edge probability $p$ depicted for small, medium and large lattices ($L = 10, 20$ and 100 respectively), depicting the phase transition between sub- and super-critical percolation at the percolation threshold $p_c$.}
	\label{fig:percolation}
	\vspace{-5pt}
\end{figure}

Commonly, schemes for generating $\mathcal{L}$ correspond to a bond-percolation, where successful bonds correspond to \emph{open} edges \cite{Grimmett1999,Stauffer1994}.
On percolated lattices with bond probability $p$, the existence of an infinite open cluster exhibits threshold behaviour.
In the limit of an infinite lattice $\mathcal{L}_\infty$, the probability $P_\infty(p, \mathcal{L}_\infty)$ that an infinite open cluster $\mathcal{C}_\infty$ undergoes a phase transition (from 0 to 1) at $p=p_c$.
This threshold represents the division between two distinct percolation regimes for $p<p_c$ and $p>p_c$, known respectively as the sub- and super-critical regime.
The degree of connectivity within the lattice is fundamentally different between these regimes; for example, the scaling in size of the largest connected component transitions from sub-linear to linear across the threshold, as depicted in Fig.\ \ref{fig:percolation} a).
For finite lattices $\mathcal{L}$, the finite-sized analogue to $P_\infty$ is probability $P_i(p, \mathcal{L})$ that a spanning cluster $\mathcal{C}$ exists along the $i$ direction, thereby containing a path connecting opposite faces of the lattice block along axis $i$. 
Thresholds for $P_i(p, \mathcal{L})$ correspond to continuous functions, becoming sharper for larger lattices and converge to $P_\infty(p, \mathcal{L}_\infty)$, as depicted by Fig.\ \ref{fig:percolation} b).
In practise, percolation thresholds can be found by identifying the crossing point of functions $P_i(p, \mathcal{L})$ for various sizes of $\mathcal{L}$ \cite{Stauffer1994}, or numerically using the Newman-Ziff algorithm \cite{Newman2001}.

In order to exploit percolation phenomena within a scheme for quantum computation, Ref.\ \cite{Gimeno-Segovia2015} also considered percolation on a subregion of the lattice with a small cross section which is to be used as a single-qubit channel for MBQC.
By simulating $P_t(p_f=0.75, \mathcal{L})$ for $\threed{L_t}{L}{L}$ brickwork diamond lattices over a range of $L$ (for $L_t \gg L$), it was shown that long-range percolation, and hence a single-qubit channel, was produced above some minimum $L$.
This result can also be applied to finding long-range renormalization.

\section{Long-range percolation for single-qubit channels} \label{sec:long-range_perc}

Our first set of new results extends the study of lattice percolation for single-qubit channels presented in Ref.\ \cite{Gimeno-Segovia2015}, which was limited to the generation of the partially-amorphous\footnote{Here \emph{partially-amorphous} describes a lattice that may contain bonds other than those defined by the lattice structure, such as diagonal edges or edges between non-adjacent nodes.
When constructing a brickwork diamond lattice by the scheme presented in Ref.\ \cite{Gimeno-Segovia2015}, this occurs for certain choices of fusion gate bases.} and anisotropic brickwork diamond lattice, built specifically with $p_f=0.75$ fusion gates.
To do so, we present a generalised model of percolation on elongated bond-percolated cubic lattices and establish a relationship between the minimum side-length $L_\emphrm{min}$ required for consistent long-range percolation and edge probability $p$.

\begin{figure}[t]
	\centering
	\vspace{0pt}
	\includegraphics[width=0.85\textwidth]{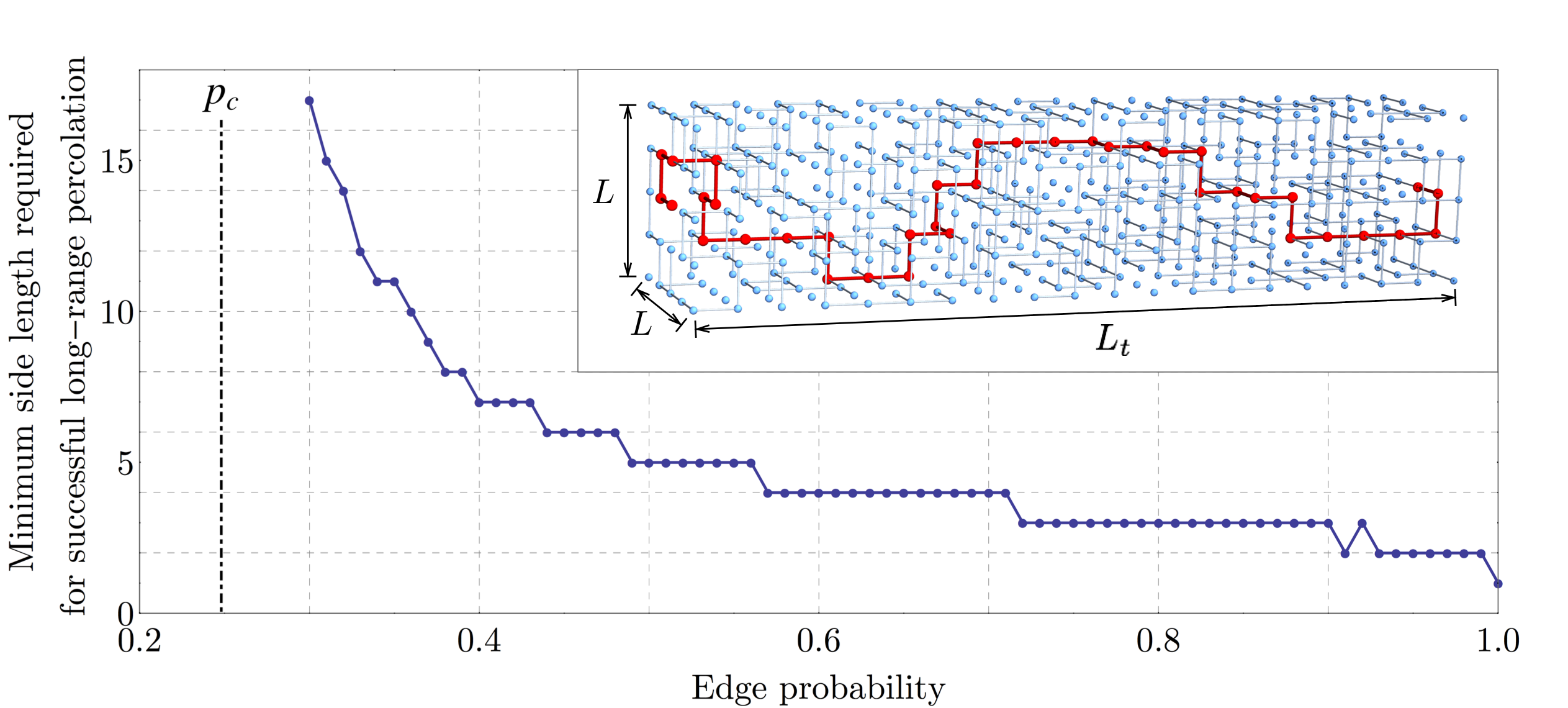}
	\caption{The minimum side length $L=L_\emphrm{min}$ required for successful long-range block percolation ($P_t(p, \mathcal{L}_t)\geq 0.95$ for $L_t=1000$) as a function edge probability $p$ for cubic lattice.
	For a given edge probability, $L_\emphrm{min}$ represents not only the smallest $L$ required for pathfinding, but also the smallest renormalization block size achievable. Inset: An illustrative example of a block of percolated cubic lattice with a valid percolated path highlighted in red.}
	\label{fig:Lmin}
	\vspace{-5pt}
\end{figure}

The model we use is as follows: consider a block of percolated $\threed{L_t}{L}{L}$ cubic lattice $\mathcal{L}_t$ with edge probability $p$, where $L_t \gg L$, depicted inset in figure 2..
On $\mathcal{L}_t$, we examine the existence of an end-to-end spanning cluster, occurring with probability $P_t(p,\mathcal{L}_t)$.
To produce a reliable single-qubit channel, we specifically consider probabilities of percolation near unity, $P_t(p,\mathcal{L}_t) \approx 1$.
We therefore generally consider \emph{successful} outcomes (for percolation and, in later sections, pathfinding) as having probability of at least 0.95, and \emph{long-range} as referring to $L_t \geq 1000$.
\textcolor{black}{These definitions are chosen such that if the above conditions are satisfied, a renormalized qubit loss rate below $10^{-3}$ can be achieved (given reasonable assumptions of renormalisation blocks with side-length $\mathcal{O}(10)$ in the scheme of Kieling, et. al. \cite{Kieling2007})}\footnote{\textcolor{black}{This can be seen by noting that if the probability of creating $100$ renormalized qubits is greater than $0.95$, then the probability of a creating a single renormalized qubit is (to a reasonable approximation) greater than $0.95^{\frac{1}{100}} \approx 0.9995$, and thus the loss rate for said qubit is less than $10^{-3}$.}}.	
\textcolor{black}{Given the known trade-off between correctability of qubit error and qubit loss for topological codes \cite{Stace2009}, minimising loss rates is essential for maximising tolerance for unavoidable computational errors. Such a low rate is also expected be a negligible contribution to renormalized qubit loss in the face of other potential sources of error within the architecture (such as photonic qubit loss, detector inefficiencies, distinguishability, etc.).}

However, within this model, percolation phenomena are less-well studied than in the standard regime.
When considering finite-sized, elongated lattices such as $\mathcal{L}_t$, it is challenging to make analytic statements about the existence of spanning clusters, as can often be done for the limit of infinite lattices.
For example, while for a lattice $\mathcal{L}_t$, one can find some $p<1$ such that $P_t(p,\mathcal{L}_t) \approx 1$, it is necessarily true\footnote{This can be seen by considering that the probability of no open edges occurring between two layers spanning the cross-section of the block is $(1-p)^{L^2}$, and hence the probability that this never occurs over $L_t$ layers is $\Gamma = (1-(1-p)^{L^2})^{L_t} \leq 1 $. Since a spanning cluster is contingent on this never occurring then $P_t(\mathcal{L})<\Gamma$, but for $p<1, L_t\rightarrow\infty \Rightarrow \Gamma\rightarrow 0$, and therefore in the limit of infinite length, percolation never occurs.} that as $L_t \rightarrow \infty$, $P_t(p,\mathcal{L}_t) \rightarrow 0$.
\textcolor{black}{As such, we highlight that all results presented in this work are expected to have some minor functional dependence on our specific definition of \emph{successful} and \emph{long-range} given above.}
Therefore, we apply a more phenomenological and empirical approach to the relevant percolation effects, and within the context of LOQC such results provide important information for designing an architecture.

We now consider the following question: What is the minimum side length $L_\emphrm{min}$ required to successfully produce a long-range spanning cluster $\mathcal{C}$ on $\mathcal{L}_t$ as a function of edge probability $p$?
To answer this question numerically, we have generated instances of $\threed{1000}{L}{L}$ sized $\mathcal{L}_t$ for a given $p$, and identified the minimum value $L=L_\emphrm{min}$ for which $P_t(p,\mathcal{L}_t) \geq 0.95$.
In Fig.\ \ref{fig:Lmin} we show values of $L_\emphrm{min}$ over a range of $p>p_c$.
We observe that for edge probabilities well above $p_c = 0.248$ (the percolation threshold for a simple cubic lattice \cite{Wang2013}), small $L_\emphrm{min}$ can be achieved (such as $L_\emphrm{min} = 5$ for $p = 0.5$), with small increases in $L_\emphrm{min}$ providing large reductions in $p$.
However, as $p$ approaches $p_c$, the scaling in $L_\emphrm{min}$ is less favourable, requiring progressively greater increases in $L_\emphrm{min}$ for incremental reductions in $p$.
This scaling region suffers from particularly punitive resource costs if used for MBQC, as the number of qubits in $\mathcal{L}_t = 1000L^2$ scales quadratically in $L$.
We also note that such a relationship for $L_\emphrm{min}(p)$ can be inverted to define $p_\emphrm{min}(L)$, such that for a given $L$, long-range percolation can only be achieved for some $p \geq p_\emphrm{min}$.

Furthermore, we can consider the implications of these results for a renormalization-based LOQC scheme.
In this context, $L_\emphrm{min}$ provides a lower bound on the side length for renormalization blocks.
Whether or not this bound can be reached depends on finding intersections between spanning clusters connecting pairs of opposing faces within a single block as well as between adjacent blocks.
This is especially problematic for $p$ close to $p_c$ as inter- and intra-block connectivity is sparse; however for $p$ well above $p_c$, the increased connectivity also increases the likelihood such intersections occur.

\section{Limited-lookahead pathfinding} \label{sec:LLAPF}

\begin{figure}[t]
	\centering
	\vspace{0pt}
	\includegraphics[width=0.7\textwidth]{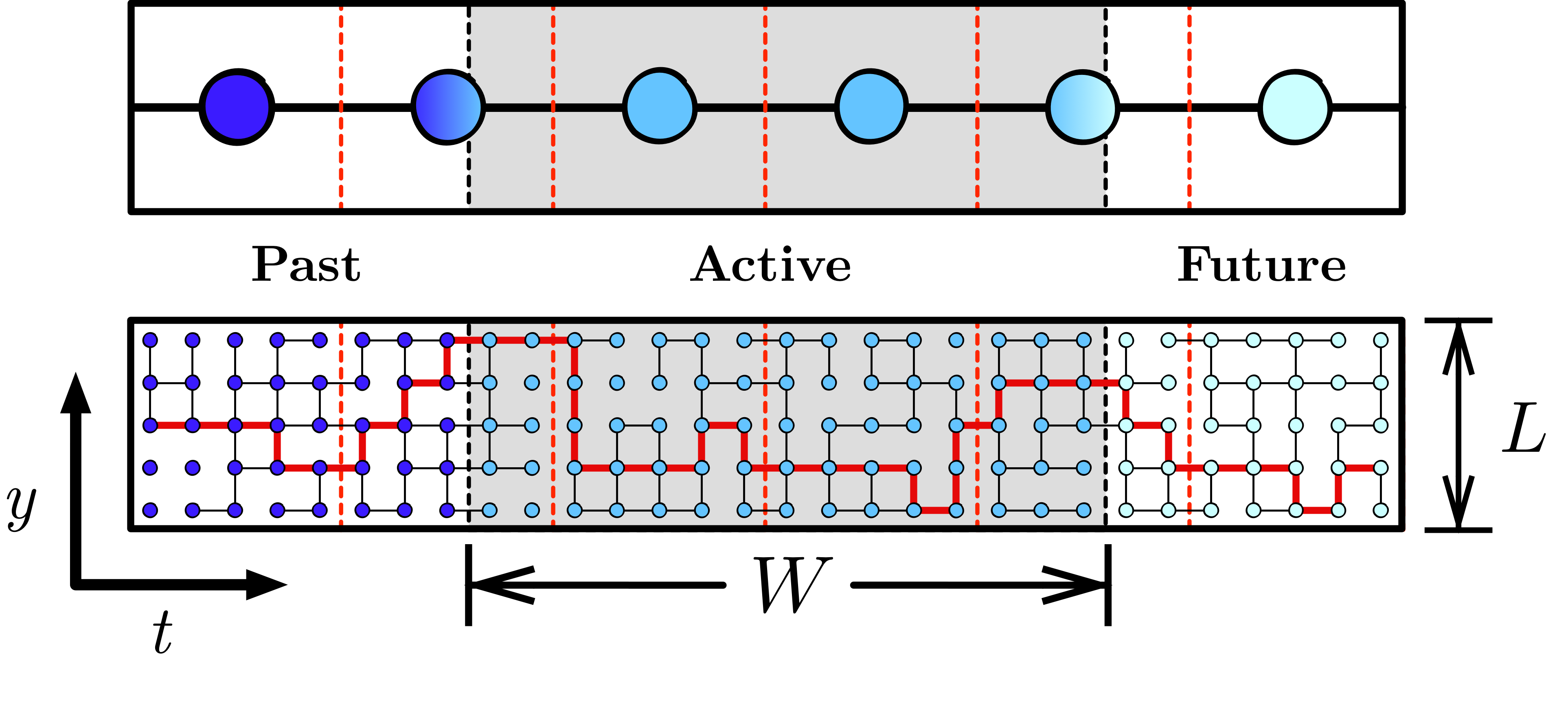}
	\vspace{-15pt}
	\caption{The renormalization process applied to a 2D lattice (existing in one time and one spatial dimension) with limited-lookahead to create a MBQC single-qubit channel.
	The lattice block can be divided into three regions in time: past, active and future.
	Past qubits exist in the past, before time $t$, having already been created and destructively measured by the device.
	Active qubits exist in the present between time $t$ and $t+W$, having been created by the device, but not yet measured.
	Future qubits exist in the future, after time $t+W$, and are yet to be created.
	Here the red, dashed lines and highlighted edges correspond to the allocation of renormalization blocks and renormalization paths respectively.}
	\label{fig:windows}
	\vspace{-5pt}
\end{figure}

In a physical LOQC device, $\mathcal{L}$ exists in one time and two spatial dimensions with $\mathbb{Z}^3$ node coordinates $(t, y, z)$ and size $\threed{L_t}{L_y}{L_z}$.
To construct $\mathcal{L}$, at each time $t$ from $t=0$ to $t=L_t$, a $\twod{L_y}{L_z}$ layer of $\mathcal{L}$ is created and entangled to the previous layer at $t-1$, where $L_y$ and $L_z$ are fixed by the renormalization protocol.
However, all-optical storage of $L_t$ lattice layers in time would require lengthy delay lines, producing a physical qubit loss rate that scales with computation length (for some applications $L_t$ is effectively unbounded); under such conditions, it is unlikely such a scheme could succeed.

It is therefore expected that an LOQC device will have a finite fixed depth, storing only a finite-depth \emph{window} $W$ of the lattice at any time $t$.
In this model, depicted in Fig.\ \ref{fig:windows}, any classical co-processing algorithms applied to $\mathcal{L}$ suffer from a \emph{limited-lookahead}, preventing analysis of a complete $\mathcal{L}$ (as previously assumed by algorithms for MBQC and renormalization).
Under this limitation, previously-considered algorithms no longer apply, or their optimality proofs and scaling efficiencies are no longer guaranteed.
To address this, new non-trivial dynamic algorithms must be designed.

However, finding optimality proofs for graph algorithms that only ever have partial knowledge of a problem is highly non-trivial, and different input scenarios may require different algorithm strategies for optimal performance.
To study the limitations of the necessary dynamic algorithms, we consider the aforementioned task of identifying single-qubit channels on percolated lattices.
Specifically, we extend the task of finding a spanning cluster presented in Sec.\ \ref{sec:long-range_perc} to the identification of a single end-to-end path, given a limited-lookahead.
To do so, we next construct a basic \emph{limited lookahead pathfinding} (LLP) algorithm.
 
\subsection{Random-node pathfinding} \label{sec:RNPF}

\begin{figure}[t]
	\centering
	\vspace{-20pt}
	\includegraphics[width=0.8\textwidth]{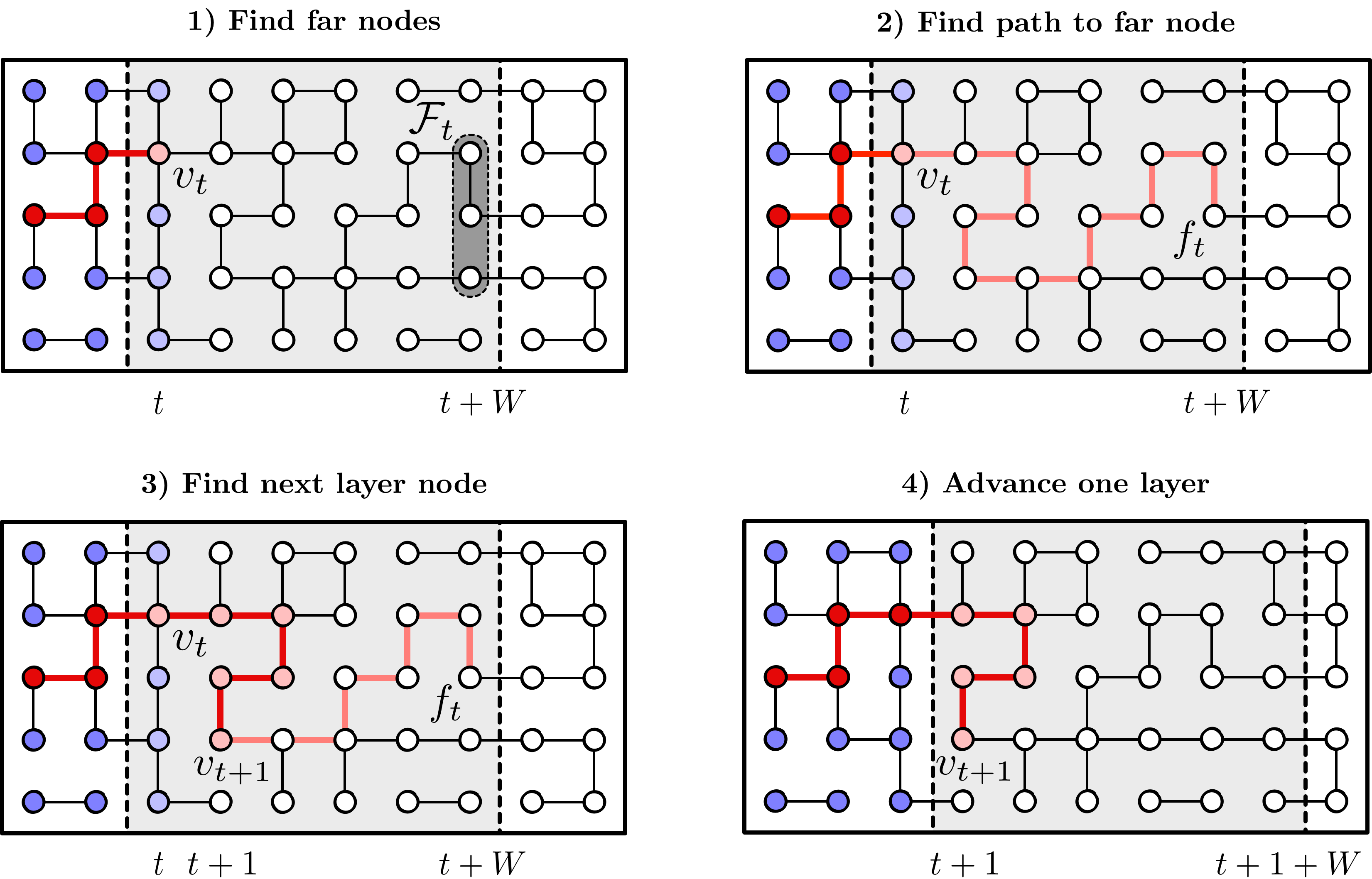}
	\caption{A single iteration of the Random-node LLP strategy with window length $W$ applied to a percolated 2D cubic lattice.}
	\label{fig:pathfinding_process}
	\vspace{-5pt}
\end{figure}

We now introduce some notation needed for describing the LLP algorithm.
Consider again the lattice $\mathcal{L}_t$ as defined in Sec.\ \ref{sec:long-range_perc}, with nodes labelled by their coordinates $(t,y,z)$.
We define a \emph{layer} $l_t$ as the subgraph of $\mathcal{L}_t$ induced by the 2D $L\!\times\! L$ layer of nodes at time $t$, that is $l_t = \mathcal{L}[\{v=(t,y,z),\, \forall\; y,z = 1,\ldots,L\}]$, \textcolor{black}{where $G'=G[V]$ denotes the \emph{induced subgraph} $G'$ of $G$ by the node set $V$.}
We define a \emph{block} $\mathcal{B}_{a,b}$ as the subgraph of $\mathcal{L}_t$ induced by the 3D block of nodes within layers $a$ to $b$ (inclusive), that is $\mathcal{B}_{a,b} = \mathcal{L}[\{v=(t,y,z),\, \forall\; t = a,\ldots,b\;,\; y,z = 1,\ldots,L\}]$.
Note that under this definition $\mathcal{L}_t = \mathcal{B}_{0,L_t}$.
The nodes within $\mathcal{B}_{a,b}$ that are also part of spanning cluster $\mathcal{C}$ of $\mathcal{L}_t$ are denoted $\mathcal{C}_{a,b} = \mathcal{C} \cap \mathcal{B}_{a,b}$ and represent nodes that are potentially usable for pathfinding.
Similarly, $\mathcal{C}_t = \mathcal{C} \cap l_t$.
In some $\mathcal{B}_{a,b}$, $\mathcal{C}_{a,b}$ may contain more than one connected component.
\textcolor{black}{Therefore, we also define $\mathcal{C}_{a,b}(v) = \mathcal{L}[\{v' \in \mathcal{C}_{a,b} \;:\; \ispath{v}{v'}\}]$ as the connected component of $\mathcal{C}_{a,b}$ containing node $v$, where $\ispath{v}{v'}$ indicates that there exists an open path connecting $v$ and $v'$. Hence, if two nodes $u$ and $v$ are not path-wise connected within $\mathcal{C}_{a,b}$, then they must exist in disjoint connected components and $\mathcal{C}_{a,b}(u) \cap \mathcal{C}_{a,b}(v) = \emptyset$}.
Lastly, the superscript $\emphrm{E-E}$ denotes components that extend \emph{end-to-end} across the layers indicated, e.g.\ $\mathcal{C}_{a,b}^{\emphrm{E-E}}$ are the components in $\mathcal{C}_{a,b}$ that have nodes in both $l_a$ and $l_b$ with the number of separate end-to-end components given by $n(\mathcal{C}_{a,b}^{\emphrm{E-E}})$.

To represent a limited lookahead, we consider the restriction that at a given time $t$, we can only have knowledge of the lattice structure within the finite block $\mathcal{B}_{t,t+W}$ of fixed \emph{window-length} $W$.
This `visible' block of lattice is known as the \emph{active block}.
At the end of every time-step, the next far layer of lattice $l_{t+1+W}$ is revealed and nearest layer layer $l_t$ is removed, the active block now becoming $\mathcal{B}_{t+1,t+W+1}$ for time $t+1$.

This limitation requires us to consider an iterative approach to finding spanning paths, which we shall call \emph{limited-lookahead pathfinding}, where each time-step the algorithm must choose a path inside the lattice based on only partial information of the lattice.
Specifically, we shall consider a low-complexity instance of pathfinding, which we call \emph{Random-node pathfinding}.
We consider a naive algorithm such as this to both identify a lower bound on the success rates of general pathfinding strategies as well as their computational complexities.
To find a path $\mathcal{P}$ the following pathfinding algorithm is applied (depicted visually in Fig.\ \ref{fig:pathfinding_process}), starting at $t=0$, (with $\mathcal{P} = {v_0}$ for some $v_0 \in \mathcal{C}_0$) and is repeated until success or failure occurs:
\begin{framed}
	\vspace{-2pt}
	\noindent
	{\bf Random-node pathfinding:}
	\begin{enumerate}[label=\textbf{\arabic*}), leftmargin=*]
	  	\setlength\itemsep{1pt}
		\item{\bf Find far nodes.} From the current path node $v_t$ in the nearest layer $l_t$, find the set of all nodes $\mathcal{F}_t=\{v \in l_{t+W} \;: \;v_t\leftrightarrow v \}$ in the farthest active block layer $l_{t+W}$ to which a path exists (only considering nodes and edges within the active block).
		If $\mathcal{F}=\emptyset$, pathfinding fails.
		\item{\bf Find path to far node.} Randomly pick a far node $f_t$ from $\mathcal{F}_t$, and find the shortest path $\mathcal{P}_t = (v_t,\ldots,f_t)$ within the window to it.
		\item{\bf Find next layer node.} Find the node in layer $l_{t+1}$ that occurs furthest along $\mathcal{P}_t$ and assign it to the next time-step path node $v_{t+1}$.
		Append the $(v_t, \ldots, v_{t+1})$ section of $\mathcal{P}_t$, to $\mathcal{P}$.
		If the final node $f_t$ in $\mathcal{P}$ is a member of $l_{L_t}$, pathfinding succeeds.
		\item{\bf Advance one layer.} Remove layer $l_t$ and reveal layer $l_{t+1+W}$.
	\end{enumerate}
	\vspace{-3pt}
\end{framed}

The first thing to note is that this algorithm is far from optimal, and in fact is almost the worst strategy one could apply (other than making deliberately bad path choices).
The only non-trivial analysis of structure occurs at step 3, where the action of finding the furthest $l_t$ layer node allows the inclusion of paths that double-back on themselves, advancing forwards and then back to layer $l_t$ before eventually reaching the final layer, an example of which is shown in step 3 of Fig.\ \ref{fig:pathfinding_process}.
The most computationally expensive operation in the algorithm occurs in step 1, when finding $\mathcal{F}_t$.
This operation consists of running Dijkstra's algorithm (for finding shortest paths on arbitrary graphs) from $v_t$, thus providing Random-node pathfinding with an overall worst-case performance of $\mathcal{O}(|E| + |V|\log|V|)$ \cite{Dijkstra1959}.

Finding optimal pathfinding strategies which demand only minimal values for $W$ is very challenging in general and the Random-node strategy can be used to explore the worst-case scenario, from which improvement may be made.
Inevitably, more complex strategies that require detailed analysis of the active block's configuration are computationally expensive, which is a major concern for real-time implementation in hardware devices.
A secondary aim of our work is therefore also to minimise the computational overhead required for pathfinding, and the Random-node strategy also adheres to this goal.

\subsection{Successful long-range pathfinding} \label{sec:LR_PF}

We now consider the conditions required for successful long-range LLP and show that these can be framed in terms of standard block percolation.
This aims to reduce the complexity of analysing a dynamic pathfinding algorithm to the simpler problem of calculating percolation statistics on small lattices.

First and foremost, pathfinding fails if no spanning cluster exists.
To ensure that a path does exist (with probability $P_t \geq 0.95$ for a given pathfinding distance $L_t$), we immediately require two conditions: $p > p_c$ and $L \geq L_\emphrm{min}(p)$.
Having satisfied these, we then seek to identify the conditions such that pathfinding almost certainly succeeds.
In this section, we prove that pathfinding always succeeds if the number of end-to-end components in each active block never exceeds one, and subsequently conjecture that successful pathfinding is only achieved if the probability of this number exceeding one is less than some small $\epsilon$.

Before outlining our argument we assert two key assumptions made.
Firstly, we assume a unique spanning cluster always exists across $\mathcal{L}$ (where \emph{unique} specifies that only one ever exists), and hence exclude any cases where long-range block percolation does not exist (e.g.\ by assuming $L > L_\textrm{min}(p)$).
The validity of this assumption is provided by recalling that for $p>p_c$ the mean size of a finite cluster decreases exponentially in $p$ \cite{Grimmett1999}, thereby preventing more than one cluster from spanning the lattice. 
Given this assumption, failure therefore only occurs from incorrect choices made during pathfinding.
Secondly, we assume that at any given time, the pathfinding algorithm may only have access to information of the lattice's structure within the active block, i.e.\ it cannot store in memory any information about past lattice structure, nor gain preemptive knowledge of any future lattice structure.
This allows us to consider each individual active block as a single instance of block percolation on a small lattice, and hence percolation statistics are constant across all active blocks.

Under these assumptions, the probability $P_{\emphrm{pf}}(\mathcal{L}_t, W)$ of pathfinding across $\mathcal{L}_t$ with window length $W$, is given by the product of the probabilities $P^t_{\emphrm{pf}}(\mathcal{B}_{t,t+W})$ that, at each time-step $t$, a path node $v_{t+1}$ in $\mathcal{B}_{t,t+W}$ is chosen that still allows for successful pathfinding to distance $L_t$, that is
\begin{align}
	P_{\emphrm{pf}}(\mathcal{L}_t,W) = \prod^{L_t-W}_{t=0} P^t_{\emphrm{pf}}(\mathcal{B}_{t,t+W}, v_t),
\end{align}
such that for $W=L_t$, $P_\emphrm{pf}(p,W) = P_\emphrm{pf}^0(\mathcal{B}_{0,L_t}) = 1$ (from our first assumption).
However, for $W<L_t$, the values of $P^t_{\emphrm{pf}}$ are less easily computed.

We can easily see that the probability of successful pathfinding given next node choice $v_{t+1}$ depends on the probability that $v_{t+1}$ exists in a component extending to the farthest layer, that is $ P^t_\emphrm{pf}(\mathcal{B}_{t,t+W} \mid v_{t+1}) = P(v_{t+1} \in \mathcal{C}^\emphrm{E-E}_{t+1,L_t})$ (where we recall that $\mathcal{C}^\emphrm{E-E}_{t+1,L_t}$ are the \emph{end-to-end} connected components contained within $\mathcal{B}_{t+1,L_t}$ that have one or more nodes in both $l_{t+1}$ and $l_{L_t}$).
However, at any given time step, we cannot know whether $v_{t+1} \in \mathcal{C}^\emphrm{E-E}_{t+1,L_t}$ or not when $t \leq L_t - W$ (by our second assumption).
Instead, we desire some active block proxy condition for $P(v_{t+1} \in \mathcal{C}^\emphrm{E-E}_{t+1,L_t})$ based only on block percolation statistics.

Specifically, we are interested here in the case of $P_{\emphrm{pf}}(\mathcal{L}_t,W) \approx 1$ and hence $P^t_{\emphrm{pf}}(\mathcal{B}_{t,t+W}) \geq 1 - \epsilon$ (where $\epsilon \ll 1$) for all $t$.
Ideally, we therefore desire some \emph{feature function} of active blocks $F: \mathcal{B}_{t,t+W} \mapsto \{0,1\}$, such that if $F(\mathcal{B}_{t,t+W}) = 1$, then $P^t_{\emphrm{pf}}(\mathcal{B}_{t,t+W})=1$ surely, but if $F(\mathcal{B}_{t,t+W}) = 0$ then $P^t_{\emphrm{pf}}(\mathcal{B}_{t,t+W})<1$.
From this, we then conjecture that if lattice parameters can be found such that $P(F(\mathcal{B}_{t,t+W}) = 1)) \geq 1 - \epsilon \; \forall\; t$, successful long-range pathfinding will be achieved.
We now prove one such feature function to be the number of end-to-end connected components within an active block, and thereby define a condition for $W$ such that $P(F(\mathcal{B}_{t,t+W}) = 1)) \geq 1 - \epsilon$.

We find that one such feature function can be defined from the uniqueness of end-to-end connected components, such that
\begin{align}
	F(\mathcal{B}_{t,t+W}) = 
 	\begin{cases}
  	\; 1    \quad \textrm{if} \quad n(\mathcal{C}^\emphrm{E-E}_{t,t+W}) = 1\\
  	\; 0	\quad \textrm{if} \quad n(\mathcal{C}^\emphrm{E-E}_{t,t+W}) > 1\\
 	\end{cases}.
\end{align}
To see that this satisfies our feature function requirements, consider two possible structures of $\mathcal{B}_{t,t+W}$, either $n(\mathcal{C}^\emphrm{E-E}_{t,t+W}) = 1$, or $n(\mathcal{C}^\emphrm{E-E}_{t,t+W}) > 1$.
In the case of $n(\mathcal{C}^\emphrm{E-E}_{t,t+W}) = 1$, the previous choice of path node was essentially irrelevant, since all connected nodes in the far layer $\mathcal{C}_{t+W}$ can be reached from any $v_t \in \mathcal{C}_t$.
If this condition is satisfied for \emph{every} active block, then at each time-step all choices of path node (using our pathfinding process) are practically equivalent, and thus $P_{\emphrm{pf}}(\mathcal{L}_t,W)=1$.
Alternatively, one can understand this by saying that if $n(\mathcal{C}^\emphrm{E-E}_{t,t+W}) = 1$, the current connected component $\mathcal{C}_{t,t+W}^\emphrm{E-E}(v_t)$ \emph{must} be part of the spanning component $\mathcal{C}_{t,L_t}$ extending to the final layer, or
\begin{align}
	\mathcal{C}_{t,t+W}^\emphrm{E-E}(v_t) \cap \mathcal{C}_{t,L_t}^\emphrm{E-E} \neq \emptyset, \label{eq:pf_succeeds}
\end{align}
assuming that $n(\mathcal{C}_{t+W,L_t}^\emphrm{E-E})=1$ (which holds for $t+W \ll L_t$, from the uniqueness of $\mathcal{C}_{0,L_t}$).

\begin{figure}[t]
	\centering
	\vspace{-20pt}
	\includegraphics[width=0.7\textwidth]{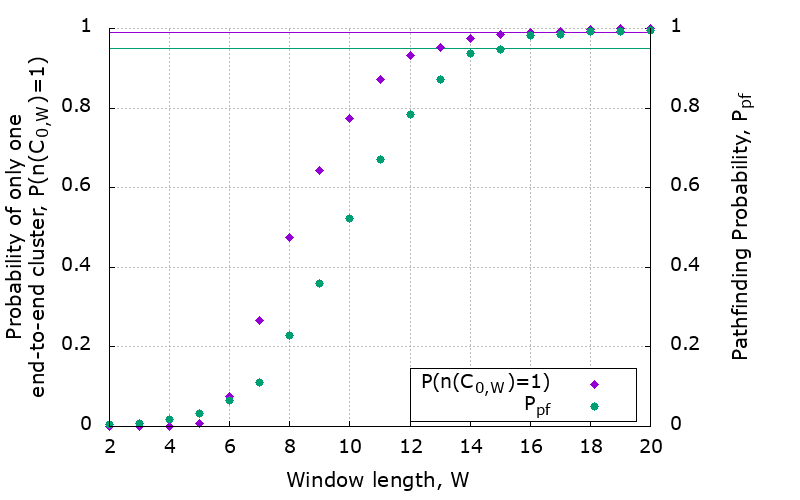}
	\caption{A comparison between $P(n(\mathcal{C}^\emphrm{E-E}_{0,W}) = 1)$ on instances of $\threed{W}{20}{20}$ edge percolated cubic lattice and the success probability of pathfinding across a lattice of size $\threed{1000}{20}{20}$ with window-length $W$ (with $p=0.3$ in both cases).
	Note that the large cross-section ($L$=20) is necessary due to $p$ close to $p_c \approx 0.248$, such that $L>L_\emphrm{min}(p)$ (see Fig.\ \ref{fig:Lmin}).
	This supports the conjecture that $P(n(\mathcal{C}^\emphrm{E-E}_{0,W}) = 1) \geq 1 - \epsilon$ is a necessary and sufficient condition for successful pathfinding, achieved for some minimum window length.
	Here we find that successful pathfinding $P_{\emphrm{pf}} \geq 95\%$ occurs for $W_\emphrm{min} \geq 16$ and is achieved for $\epsilon \leq 0.01$, with both thresholds respectively depicted by coloured lines.}
	\label{fig:cubic_perc_and_pathf}
	\vspace{-10pt}
\end{figure}

Conversely, if $n(\mathcal{C}^\emphrm{E-E}_{t,t+W}) > 1$, no choice of $v_t \in \mathcal{C}_t$ can possibly allow for all nodes in $\mathcal{C}_{t+W}$ to be reached, and hence presents a possibility that $v_t$ is not in a component that extends forward to the final layer, $v_t \notin \mathcal{C}^\emphrm{E-E}_{t,L_t}$.
In such a scenario, two possibilities exist: either, equation~(\ref{eq:pf_succeeds}) is ultimately satisfied, indicating that a path passing through $\mathcal{C}_{t,t+W}(v_t)$ can reach layer $L_t$, and therefore allows successful pathfinding, or
\begin{align}
	\mathcal{C}_{t,t+W}^\emphrm{E-E}(v_t) \cap \mathcal{C}_{t,L_t}^\emphrm{E-E} = \emptyset \label{eq:pf_fails},
\end{align} 
indicating that structure within $\mathcal{C}_{t,t+W}(v_t)$ cannot contribute to pathfinding, and therefore represents a dead-end, which causes pathfinding to fail.
Note, due to effect of finite block side lengths $L$, there is always some non-zero probability that equation~(\ref{eq:pf_fails}) is satisfied (such as no open edges existing between nodes in $\mathcal{C}_{t+W}(v_t)$ and $\mathcal{C}_{t+W+1}$) and thus $P_{\emphrm{pf}}(\mathcal{L}_t,W)<1$.
This proves that the condition $n(\mathcal{C}^\emphrm{E-E}_{t,t+W})=1$ satisfies our desired feature function requirements.

We now consider the lattice requirements such that $P(n(\mathcal{C}^\emphrm{E-E}_{t,t+W})=1) \geq 1 - \epsilon \; \forall\; t$.
To satisfy this requirement, we define (for a given $p$ and $L$) the \emph{minimum window length} $W_\emphrm{min}(L,p)$ as the smallest $W$ such that $P(n(\mathcal{C}^\emphrm{E-E}_{t,t+W}) = 1) \geq 1 - \epsilon \; \forall\; t$.
Note that for $L \geq L_\emphrm{min}$ such a minimum window length must exist; for any $\mathcal{L}_t$, clearly $P(n(\mathcal{C}^\emphrm{E-E}_{0,W}) = 1) = 1$ when $W=L_t$ (from the uniqueness of $\mathcal{C}^\emphrm{E-E}_{0,W}$), but as $W$ is decreased from $L_t$, either $W_\emphrm{min}$ is found when $P(n(\mathcal{C}^\emphrm{E-E}_{t,t+W}) = 1) < 1 - \epsilon$ occurs, or else no lookahead is required (and $W_\emphrm{min} = 2$).
We further define $W_\emphrm{max}(L,p_\emphrm{min})$ as the maximum window length required by LLP occurring for a given $L$ at $p_\emphrm{min}$, above which any further increase in $W$ provides no advantage.

We have shown that for a given $\mathcal{L}_t$ with lattice parameters $p$ and $L$, $n(\mathcal{C}^\emphrm{E-E}_{t,t+W}) = 1$ is a sufficient feature function.
We hence conjecture that $P(n(\mathcal{C}^\emphrm{E-E}_{0,W}) = 1) \geq 1 - \epsilon$ is a necessary and sufficient condition for successful long-range LLP. That is, if this condition is not satisfied then no strategy (regardless of complexity) can ever produce successful long-range LLP, and that this condition is always satisfied for $W \geq W_\emphrm{min}$ as $\epsilon \rightarrow 0$.

\subsection{Numerical simulation} \label{sec:pathf_sim}

\begin{figure}[t]
	\centering
	\vspace{0pt}
	\includegraphics[width=0.6\textwidth]{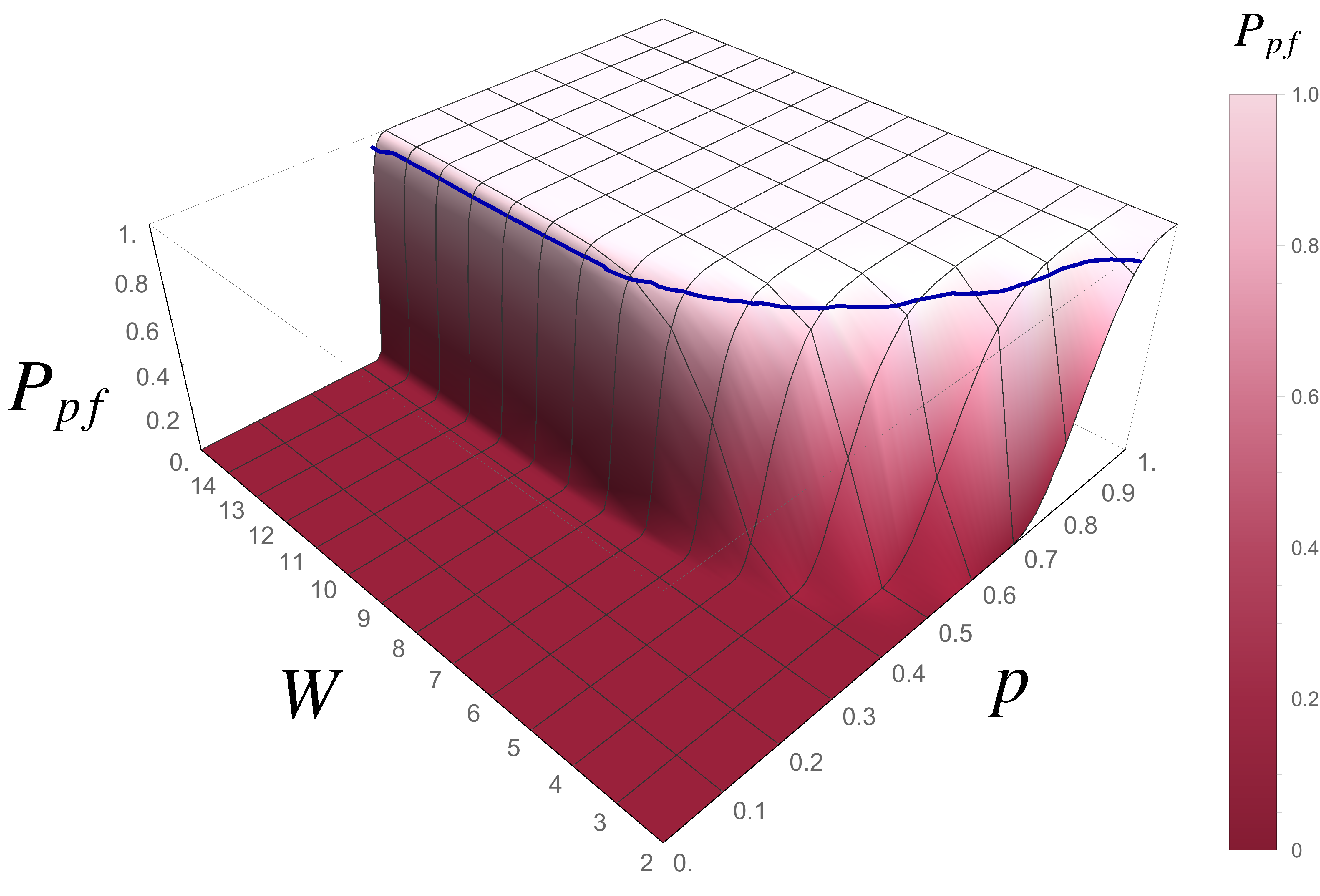}
	\caption{Limited-lookahead pathfinding success probability $P_\emphrm{pf}(p,W)$ for $\threed{1000}{7}{7}$ cubic lattices found over a range of window sizes $W$ and edge probabilities $p$.
	These results clearly depict the combination of both long-range block percolation phenomena and the effect of a limited lookahead on pathfinding.
	Firstly, a clear percolation threshold is observed at $p_\emphrm{min} \approx 0.4$, as predicted (by numeric simulations for $L_\emphrm{min}$ in Fig.\ \ref{fig:Lmin}).
	Secondly, the detrimental effect of a limited lookahead on pathfinding for window sizes $W < 10$ is also observed.
	This shows that for $p=p_\emphrm{min}$, a maximum window length $W_\emphrm{max}(L,p_\emphrm{min})$ exists, below which pathfinding can only be achieved by a complementary increase in $p$.
	The region of successful long-range pathfinding ($P_\emphrm{pf}(p,W) \geq 0.95$) is found above the highlighted blue contour.}
	\label{fig:cubic_percolation_7x7}
	\vspace{-5pt}
\end{figure}

We now consider numerical simulation of LLP applying a Random-node strategy.

Firstly, we address the conjecture that $P(n(\mathcal{C}^\emphrm{E-E}_{0,W})=1) \geq 1 - \epsilon$ is a necessary and sufficient condition for successful pathfinding.
Fig.\ \ref{fig:cubic_perc_and_pathf} depicts simulation of both LLP and $\mathcal{B}_{0,W}$ block percolation over a range of $W$ for a cubic lattice.
We observe that successful pathfinding occurs for minimum window length $W_\emphrm{min}(20,0.3)=16$, where $P_\emphrm{pf}(p,W) = 0.983$ and $P(n(\mathcal{C}^\emphrm{E-E}_{0,W})=1) = 0.991$, such that $\epsilon = 10^{-2}$.
We also note that $P_\emphrm{pf}(p,W)$ drops significantly as $P(n(\mathcal{C}^\emphrm{E-E}_{0,W})=1)$ decreases below $1 - \epsilon$, further validating our choice of feature function.
In conjunction with the proofs of our feature function presented in Sec.\ \ref{sec:LR_PF}, these results support our conjecture that $P(n(\mathcal{C}^\emphrm{E-E}_{0,W})=1) \geq 1 - \epsilon$ is a necessary and sufficient condition for successful pathfinding.

We now consider the interdependence of pathfinding parameter $W_\emphrm{min}$ and lattice parameters $L$ and $p$.
To do so, we consider the probability of successful pathfinding $P_\emphrm{pf}(p,W)$ on instances of cubic lattice $\mathcal{L}_t$ with dimensions $\threed{1000}{L}{L}$ over a range of $p$ and $W$.
Fig.\ \ref{fig:cubic_percolation_7x7} depicts such a simulation for $L=7$.

\begin{figure}[t]
	\centering
	\includegraphics[width=0.85\textwidth]{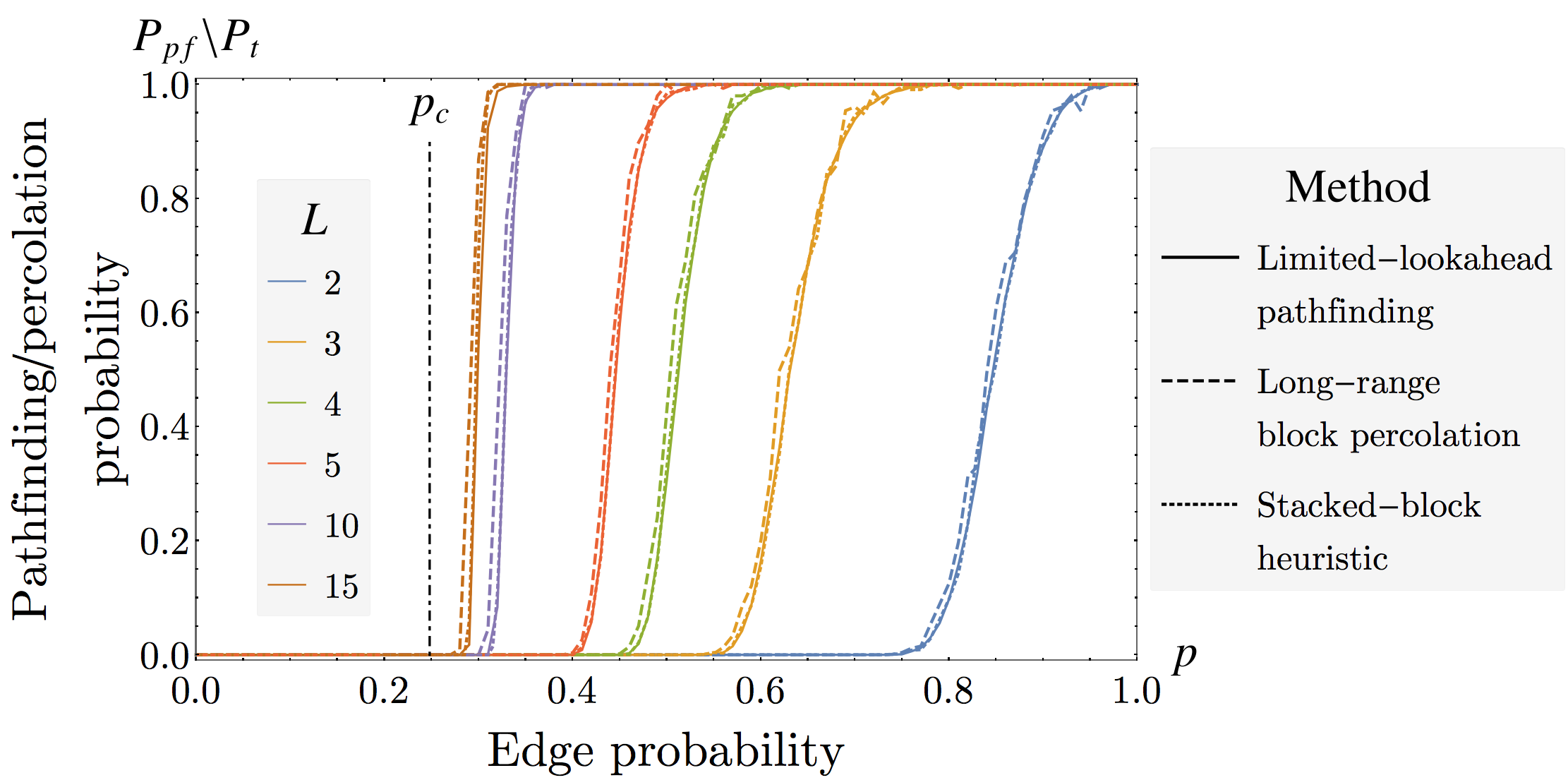}
	\caption{Comparison between the thresholds in LLP success rates $P_\emphrm{pf}$ and block percolation $P_t$ on instances $\threed{1000}{L}{L}$ cubic lattices over a range of $L$.
	For pathfinding, depicted by solids lines, a large window size (of $W=15 \geq W_\emphrm{max}(L)$) was chosen to ensure the thresholds found were due to percolation effects, rather than pathfinding's limited lookahead.
	By comparison with long-range block percolation, depicted by dashed lines, we can see that $P_\emphrm{pf} \approx P_t$, confirming that for sufficiently large window lengths, LLP is equivalent to long-range block percolation.
	Furthermore, we find that within this regime both long-range block percolation and LLP can be approximated as multiple stacked instances of ($\threed{15}{L}{L}$) active block percolation, such that $P_{\emphrm{pf}}(p,\mathcal{L}_t,15) \approx P_t(p,\mathcal{L}_t) \approx P_t(p,\mathcal{B}_{0,15})^\frac{1000}{15}$, as depicted by the dotted lines.
	Given that simulating LLP is computationally expensive, this \emph{stacked-block} heuristic provides a quick and inexpensive approximation for investigating the performance of LLP on other percolated lattices for LOQC.}
	\label{fig:perc_and_pathf_thresholds}
	\vspace{-5pt}
\end{figure}

The first and most striking feature of these results is the sharp threshold at $p \approx 0.4$ for large $W$.
This clearly identifies the minimum edge probability $p_\emphrm{min}(L=7)$ below which no long-range percolation occurs, and agrees with numerical $L_\emphrm{min}$ results depicted in Fig.\ \ref{fig:Lmin}, showing that $p_\emphrm{min}(L=7) \approx 0.4$.
From the argument made in Sec.\ \ref{sec:LR_PF}, we expect this pathfinding threshold to recreate the standard block percolation threshold of a $\threed{1000}{L}{L}$ cubic lattice.
We confirm this numerically with Fig.\ \ref{fig:perc_and_pathf_thresholds}, which depicts LLP and block percolation thresholds found over a range of $L$, showing LLP reproducing long-range block percolation statistics.
Furthermore, we find that percolation statistics found for active blocks can be used to estimate pathfinding performance over long distances.
In this simplified \emph{stacked-block} heuristic, we model long-range LLP as $1000/W$ consecutive instances of block percolation, as if adjacently stacked face-to-face in $t$ to form the full block $\mathcal{L}_t$ (without requiring two adjacent blocks' percolation paths are connected at adjacent faces), such that $P_{\emphrm{pf}}(p,W) \approx P_t(p,\mathcal{B}_{0,W})^\frac{1000}{W}$.
Fig.\ \ref{fig:perc_and_pathf_thresholds} shows that even for large $L$, this heuristic provides a good estimate for $P_\emphrm{pf}(p, W)$ and $P_t(p)$ (when $W \geq W_\emphrm{max}$).

The second feature we observe is the effect of small window lengths upon pathfinding.
For $p = p_\emphrm{min}$, we observe a maximum window length $W_\emphrm{max}(L,p_\emphrm{min}) \approx 10$.
As conjectured, $W > W_\emphrm{max}$ provides no additional benefit to pathfinding, whereas for $W < W_\emphrm{max}$, the probability of successful LLP is significantly reduced (for fixed $p$).
While it is possible to realise successful pathfinding for $W < W_\emphrm{max}(L,p_\emphrm{min})$, this can only be achieved by a complementary increase in $p$.

\begin{figure}[t]
	\centering
	\vspace{0pt}
	\includegraphics[width=0.7\textwidth]{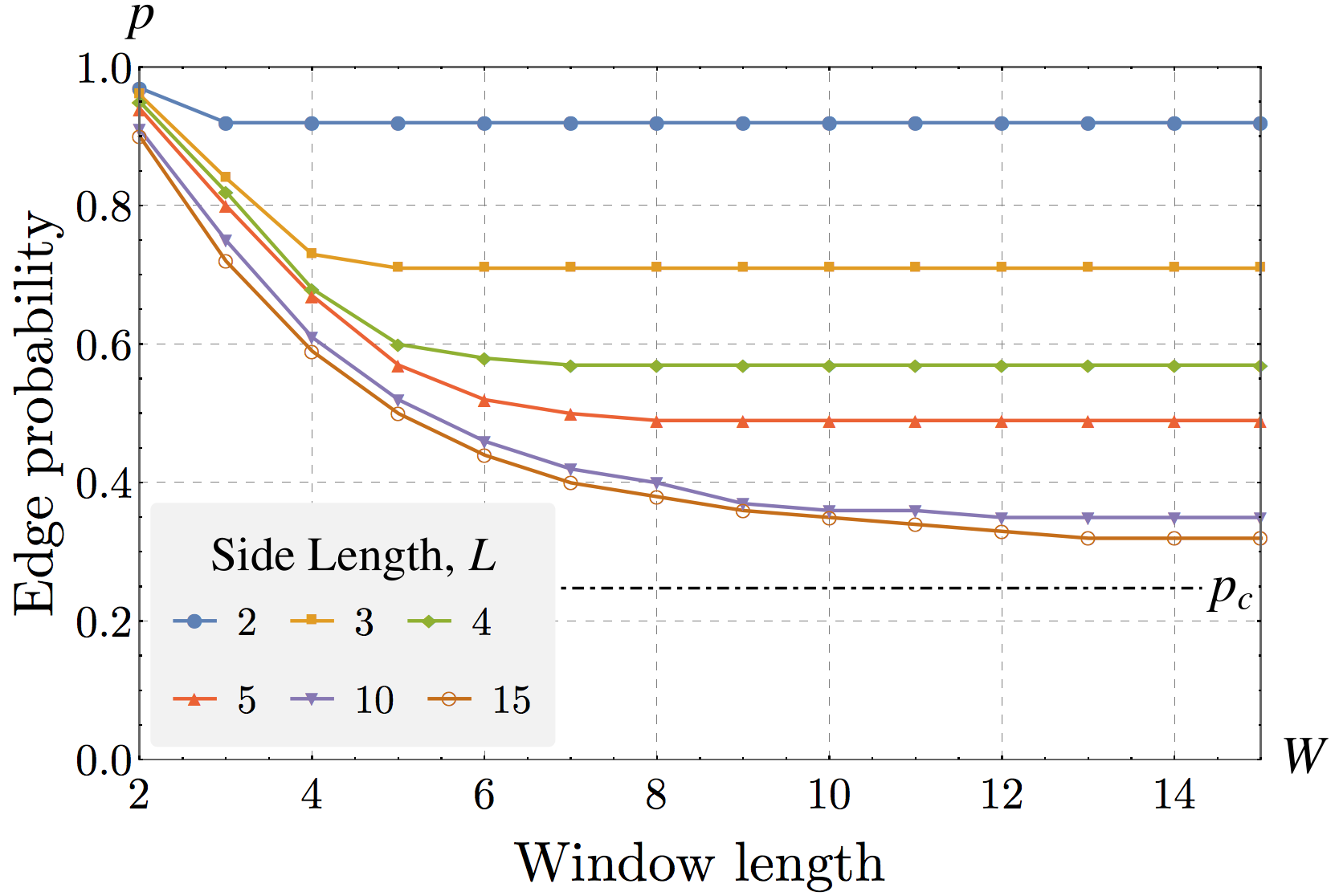}
	\caption{Contours of successful pathfinding ($P_\emphrm{pf}(p, W) = 0.95$ for $\mathcal{L}_t$ with dimension $\threed{1000}{L}{L}$) for a range of side lengths $L$.
	From this, we can fully understand the various resource trade-offs one can make in order to achieve successful pathfinding.}
	\label{fig:cubic_pathfinding_contour}
	\vspace{-5pt}
\end{figure}

To fully understand the parameter space for successful pathfinding, we consider contours of $P_\emphrm{pf} = 0.95$ in $p$ and $W$ for $L = 2,3,4,5,10$ and $15$, depicted in Fig.\ \ref{fig:cubic_pathfinding_contour}.
From these results we can also incorporate the effects of $L$ into our previous analysis.
As identified by the results of Fig.\ \ref{fig:Lmin}, an increase in $L$ reduces the minimum edge probability $p_\emphrm{min}$ at which long-range percolation occurs, and hence the value of $p_\emphrm{min}$ which LLP can succeed.
However, whilst an increase in $L$ (for a fixed $W$) always decreases the required $p$ for successful pathfinding, these gains are most significant when $W$ is also increased, allowing the new $p_\emphrm{min}(L)$ to be achieved.
Such insights provide us with far greater clarity into the inherent resource trade-offs in a LOQC device.

Finally, we note that even for the largest active blocks considered, $p_\emphrm{min}(L=W=15)$ had yet to approach $p_c$.
This indicates that successful pathfinding is likely to require a lattice with edge probability greater than $p_c$ by some non-insignificant amount.
Furthermore, when more sophisticated and computationally expensive pathfinding strategies were simulated, they did not reduce $W_\emphrm{max}(L,p_\emphrm{min})$, only improving pathfinding in the region of $p>p_\emphrm{min}$ and $W<W_\emphrm{max}(L,p_\emphrm{min})$. 
For further details on the performance of such strategies, see the Supplementary Materials.

\section{Implications for LOQC architectures} \label{sec:implications}

Using the results presented in Sec.\ \ref{sec:pathf_sim} additional clarity can now be given to the resource trade-offs inherent to a realistic LOQC device.

Firstly, generating a lattice with $p>p_c$ is necessary for the reduction of active block size.
For $p$ close to $p_c$, small increases in $p$ will lead to significant resource savings in block size.
The success rate of LOQC's boosted fusion gates\footnote{Note $p \neq p_f$, as in current proposals multiple fusion operations must succeed for a given edge to be created in the target lattice.
Furthermore, failure modes of boosted fusion gates can also maintain connectivity, producing additional connectivity outside the standard percolation model.} $p_f$ can be increased from 50\% to 75\% through the consumption of either a Bell state or four single photons per gate \cite{Grice2011,Ewert2014}.
However, above this first level of boosting, gains in $p_f$ become more marginal at the expense of increasingly costly resource states (which cannot be produced deterministically using linear optics without significant resource overheads).
This leads us to believe that it is likely that LOQC will utilise boosted fusion of at least $p_f=75\%$, from which a choice of active block dimensions, $W$ and $L$, can be made accordingly.
We note that in Ref.\ \cite{Gimeno-Segovia2015}, it was shown that $p_f = 75\%$ greatly exceeds the percolation threshold of $p_c = 62.5\%$.
In practise, experimental fusion gate success rates will be reduced by error mechanisms, such as photon loss.
However, if this reduction can be sufficiently minimised, our results indicate that small active block sizes can be achieved, thereby reducing overall resource requirements for LOQC.

Secondly, the probability of successful pathfinding affects the accommodation of bond/qubit\footnote{If a block lacks connectivity to be successfully renormalized, one can choose to represent this either as the loss of individual bonds or an entire qubit.} loss for a renormalized lattice.
From the perspective of the lattice renormalization, a failure in pathfinding simply represents a missing bond/qubit along the time axis.
Thus the quantum error correction (QEC) protocol's ability to deal with bond/qubit loss on the renormalized lattice explicitly determines the required $P_\emphrm{pf}$ (which adds to all other loss mechanisms).
For example, consider the pathfinding requirements for a linear cluster of 100 renormalized qubits, with each renormalized block being 10 layers long, such that the dimensions of $\mathcal{L}_t$ are $\threed{1000}{L}{L}$.
If less that one bond/qubit must be lost per string of 100 renormalized qubits, then we require $P_\emphrm{pf}(\mathcal{L}_t,W)>0.99$.
However, if more bond/qubit loss can be accommodated, this reduces the required pathfinding probability, thus allowing for a further reduction in $L$ or $W$.

\textcolor{black}{Finally, we expect the identified resource costs and trade-offs to be somewhat sensitive to our chosen value of $P_\emphrm{pf}$, and would expect a reduction in size of the successful long-range pathfinding parameter space ($L$, $W$, $p$) if it were increased (say to 0.99).}
\textcolor{black}{However, we further expect that the effect of such a difference would be very small and furthermore would decrease}\footnote{\textcolor{black}{This can be understood by consideration of Figure 6. Here we can observe that an increase of $P_\emphrm{pf}$ from 0.95 to 0.99 only provides a small contraction of the space outlined by the highlighted contour, a difference which clearly decreases as $P_\emphrm{pf} \rightarrow 1$.}} \textcolor{black}{as $P_\emphrm{pf} \rightarrow 1$, and therefore our presented results provide an accurate description of the relevant limited-lookahead phenomenon.}

\section{Open questions} \label{sec:open_questions}

There are other architectural necessities that must be incorporated to produce a complete model.
In this work pathfinding is only considered within the context of producing a single-qubit channel, but in order to produce a renormalized lattice for QEC percolated paths must also be found in $y$ and $z$.
While an renormalization algorithm with optimal scaling is known for 2D \cite{Browne2008}, none are known for higher-dimension lattices.
Additionally, for a realistic device, local pathfinding algorithms must also be designed to reduce the associated computational overheads for finding percolated paths in both $y$ and $z$ (for example, similar to recently proposed cellular automata decoders for QEC \cite{Herold2014a}).

Also, we do not consider the effects of experimental errors on our pathfinding strategy.
It is known that one of the most significant challenges for LOQC is photon loss.
The teleportation of quantum information via MBQC in our model assumes that each photon is measured successfully.
However, in a physical device some degree of both heralded and unheralded photon loss will undoubtably occur from  active components and memory delay lines.
For heralded qubit loss occurring in the lattice generation stage, it is known that the affected qubit's neighbours can be removed from the lattice.
With this approach, it was shown in Ref.\ \cite{Gimeno-Segovia2015} that a loss rate up to 1.5\% could be tolerated by the diamond brickwork lattice (with $p_f = 75\%$).
Given that for $W\geq W_\emphrm{max}(L,p)$ we recover standard percolation statistics, we therefore expect a similar loss tolerance for our pathfinding model.
But for an unheralded qubit loss it is not yet known whether it is possible to perform MBQC without an explicit loss-tolerant encoding (such as presented in Ref.\ \cite{Varnava2006}), especially under the realistic restriction of a fixed order of qubit measurement.

We additionally note that in the context of a LOQC architecture, our approach here is far from optimal.
For example, our pathfinding algorithm only considers a single path per qubit channel at anyone time.
However, for $p \gg p_c$ the number of percolation paths spanning one axis of a $\threed{L}{L}{L}$ block scales as $\mathcal{O}(L)$, compared to $\mathcal{O}(1)$ for $p>p_c$ close to $p_c$ \cite{Franceschetti2007}.
It may therefore be possible to utilise these extra paths as \emph{backup} paths to insure against both unheralded photon loss and unforeseen dead ends.
This may have the combined effect of both reducing $W_\emphrm{min}(L,p)$ and providing loss tolerance, without resulting in an increased susceptibility to accrued Pauli errors (from increased MBQC measurements per single-qubit channel).

\textcolor{black}{Lastly, it remains to extend such pathfinding simulations to candidate lattices for percolated LOQC cluster states. Due to the amorphism, anisotropy and correlations of bond percolation applied to the brickwork diamond lattice presented in Ref.\ \cite{Gimeno-Segovia2015}, a direct mapping of resource costs cannot be made from our results. However, preliminary simulations have shown comparable effects as presented here, suggesting that the presented LLP phenomenon is general to many lattice configurations \cite{Morley-Short18}. Nevertheless, it remains to identify the specific impact of deviations from the standard percolation model as such lattices must also permit resource-efficient LLP in order to be utilised within an LOQC architecture.}

\section{Conclusions and Outlook}

Realistic architectures for LOQC must consider the physical constraints of a large-scale device, such as a finite and fixed depth.
As such, this work has considered the effect of a finite fixed depth on the creation of a single-qubit channel from a percolated cluster state lattice.
We have shown that within this model, a limited-lookahead pathfinding algorithm can be applied to successfully create such a channel and identified resources requirements for successful pathfinding.
\textcolor{black}{This suggests that an LOQC architecture with a computational window of $\mathcal{O}(10)$ layers (i.e. clock-cycles of photon production) is sufficient to produce the almost indefinitely large states required for universal quantum computation.}
However, we also find that these constraints many require percolation-based LOQC architectures to operate above previously-identified minimum resource estimates.
Notably, we find that resource requirements become significant as the cluster state lattice's edge probability approaches it's critical threshold.
However, this equally implies that even small increases in edge probability (close above the percolation threshold) can provide significant resource savings and allow an LOQC device to operate with surprisingly low fixed depth.

An additional key result of this work is a significant step towards bridging the gap between high- and low-level architectural requirements. 
When applied to a specific LOQC architectural schema, the model presented here allows direct mapping of high-level architectural resource requirements (such as a qubit channel loss rates) onto low-level device requirements (such as device depth and ancillae resource counts).
Once identified, this mapping allows the device's fixed finite depth to be effectively ignored allowing the high-level abstractions required for studying the high-level architecture, such as QEC protocols.
Furthermore, by identifying LLP simulation heuristics, the performance of novel candidate lattices for LOQC can be quickly and easily analysed without extensive LLP simulations---a key advantage as architectural models become increasingly sophisticated.

\printbibliography

\section*{Acknowledgements}
\textcolor{black}{The authors thank an anonymous referee for suggesting an additional pathfinding strategy included in the Supplementary Materials.}
This work was supported by the UK Engineering and Physical Sciences Research Council (EPSRC).
SMS is supported by the Bristol Quantum Engineering Centre for Doctoral Training, EPSRC grant EP/L015730/1.
MGS acknowledges support from the University of Bristol.
This work was carried out using the computational facilities of the Advanced Computing Research Centre, University of Bristol.
\textcolor{black}{All simulation and analysis scripts and datasets presented here are available for download from the Research Data Repository of University of Bristol at \url{data.bris.ac.uk/data/dataset/2wmj58va0tejt23ojmkxlkf4nu}.}

\newpage

\section*{Supplementary materials}

\subsection*{Other pathfinding strategies}

By considering other strategies for LLP, we now present further evidence to support the conjecture of Sec.\ \ref{sec:LR_PF}, which states that: ``if [$P(n(\mathcal{C}^\emphrm{E-E}_{0,W})=1) \geq 1 - \epsilon$] is not satisfied then no strategy (regardless of complexity) can ever produce successful long-range LLP''. 
\textcolor{black}{Addressing the validity of such a statement is equivalent to the answering the question: ``does an algorithm exist that can achieve successful long-range LLP on active blocks that contain more than one end-to-end connected component (with some probability greater than $\epsilon$)?''. We provide the following results to suggest that the answer this question is ``No.''.}

All results previously presented utilised the Random-node strategy, where the choice of node in the active block's farthest layer (to find a path to) was made at random.
We now introduce three variants of the Random-node strategy, all providing a different metric for far-layer node choice:
\begin{itemize}
	\item{\bf Shortest-path:} pick the node in the farthest layer to which the shortest path exists.
	\item{\bf Most-connected:} pick the node in the farthest layer with the highest degree.
	\item{\bf Centre-first:} pick the node in the farthest layer that is most central in the $y$--$z$ plane. 
\end{itemize}
In all strategies, if multiple nodes are found as equal best choice, one is selected at random. 
Also as before, once a far node has been selected the shortest path to it is always found.

\textcolor{black}{We additionally present a pathfinding algorithm of increased complexity, named the ``Most-paths'' strategy. In this algorithm, the next node is found by identifying which node in the next-nearest layer has paths to the greatest number of nodes in the far layer. Such a strategy thereby requires $\mathcal{O}(L^2)$ applications of Dijkstra's algorithm per time step, as opposed to the single use demanded by Random-node pathfinding (and it's variants).}
 
\begin{figure}[b!]
	\centering
	\vspace{-5pt}
	\includegraphics[width=0.7\textwidth]{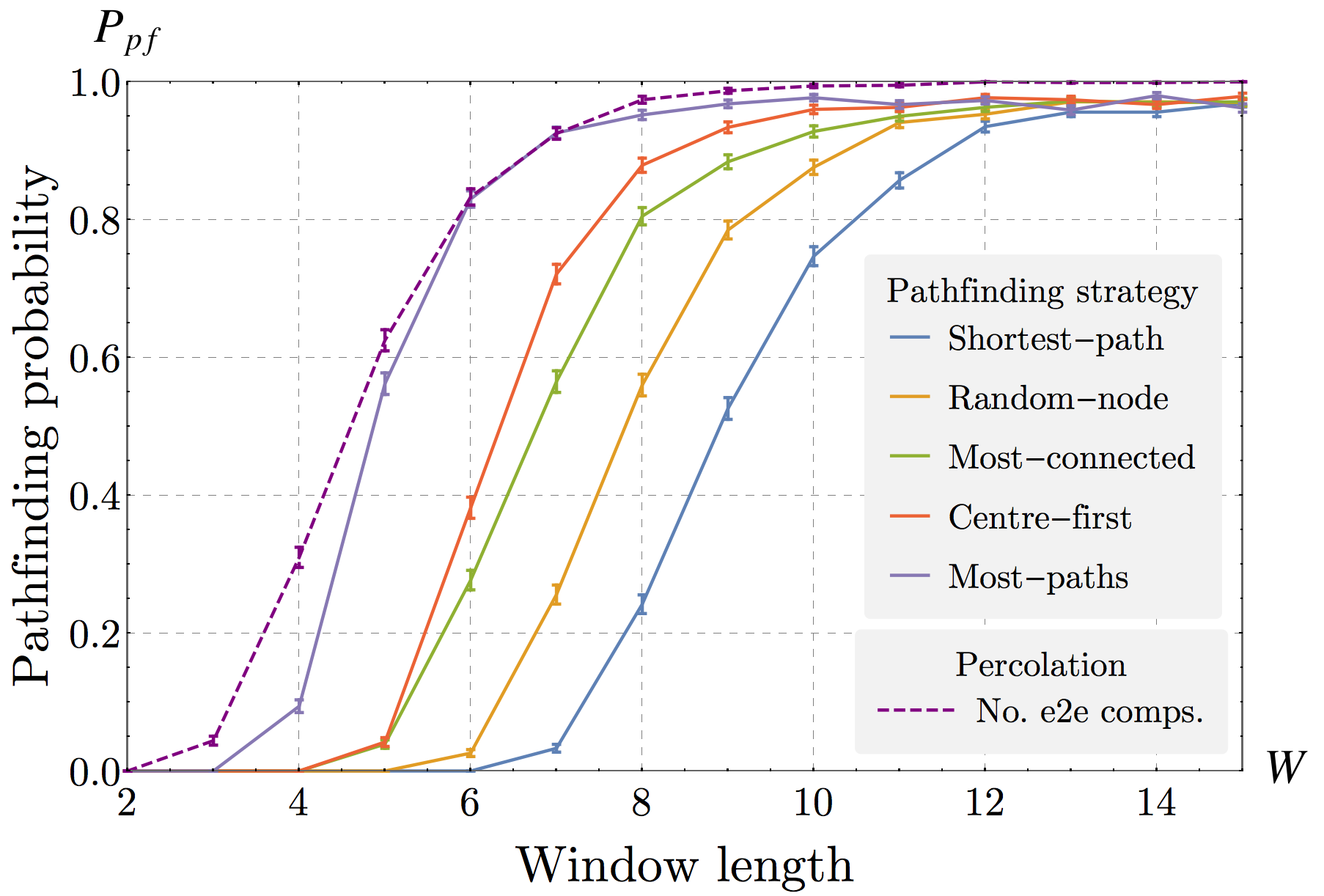}
	\vspace{-5pt}
	\caption{The performance of LLP strategies on $p=0.35$ instances of $\threed{1000}{10}{10}$ cubic lattice over a range of window lengths $W$. 
	\textcolor{black}{We find that no strategy achieves LLP for $\epsilon > 0.025$ which is tightened to $\epsilon > 0.006$ for low-complexity strategies.}
	These results also suggest that a realistic LOQC device will not require complex LLP strategies to achieve near-perfect LLP, thereby reducing loss-rates inflicted by photon delay lines.} 	
	\label{fig:cubic_pathfinding_strategies}
	\vspace{-10pt}
\end{figure}

The performances of the all five strategies for LLP are shown in Fig.\ \ref{fig:cubic_pathfinding_strategies} depicted for $p=0.35$ instances of a $\threed{1000}{10}{10}$ cubic lattice over a range of $W$.
\textcolor{black}{We find that no strategy achieves successful long-range LLP for $\epsilon > 0.026$, with best performance achieved by the Most-paths strategy with $W = 8$. For the lower-complexity Random-node variants, no strategy succeeds for $\epsilon > 0.006$, with the best performance achieved by a Centre-first strategy with $W = 10$. Such low bounds on $\epsilon$ clearly support our conjecture. Interestingly, we note that the original Random-node strategy does not provide the lowest performance, outperforming Shortest-path. This counter-intuitive result highlights the difficulty in designing effective pathfinding algorithms as well as analysing the causes of their success/failure.}

\textcolor{black}{These results highlight an additional trade-off within the LOQC architecture between the device's  physical depth $W$ and the length of delay-line needed for any classical co-processing time. While the performance of the Most-paths strategy suggests that successful pathfinding can be achieved for greater values of $\epsilon$ than expected, such an improvement only allows the reduction of $W$ by 2 or 3 layers. Given that this reduction comes at a cost of $\mathcal{O}(L^2) \approx 100$ times more classical co-processing per pathfinding time-step, it is unlikely that such a trade-off would be desired. This can be seen by noting that the total delay-time $\tau_\emphrm{delay}$ demanded for current LOQC architectures can be approximately given by $\tau_\emphrm{delay} = W\tau_\emphrm{LLP}$ where $\tau_\emphrm{LLP}$ is the worst-case time taken for the classical co-processing of LLP. Given that photon loss is exponential in $\tau_\emphrm{delay}$, clearly, any reduction in $W$ must not be offset by any subsequent increase in $\tau_\emphrm{LLP}$. We therefore expect that an LOQC architecture is likely to utilise a low-complexity LLP algorithm, such as the Random-node variants considered here.}

\subsection*{Further numeric analysis}

Here we present numerical results comparing percolation and pathfinding statistics to further explore their explicit dependance.
Ideally, it would be desirable to have a quantitative relationship between rates of successful LLP and percolation statistics, such that $P_\emphrm{pf} \approx f(P(n(\mathcal{C}^\emphrm{E-E}_{0,W})=1))$ for some percolation to pathfinding rate conversion function $f$.
If a suitable $f$ can be found, this would significantly improve our theoretical understanding of LLP dynamics as well as providing more robust heuristic methods for analysing novel architectures.

\begin{figure}[t]
	\centering
	\vspace{0pt}
	\includegraphics[width=0.7\textwidth]{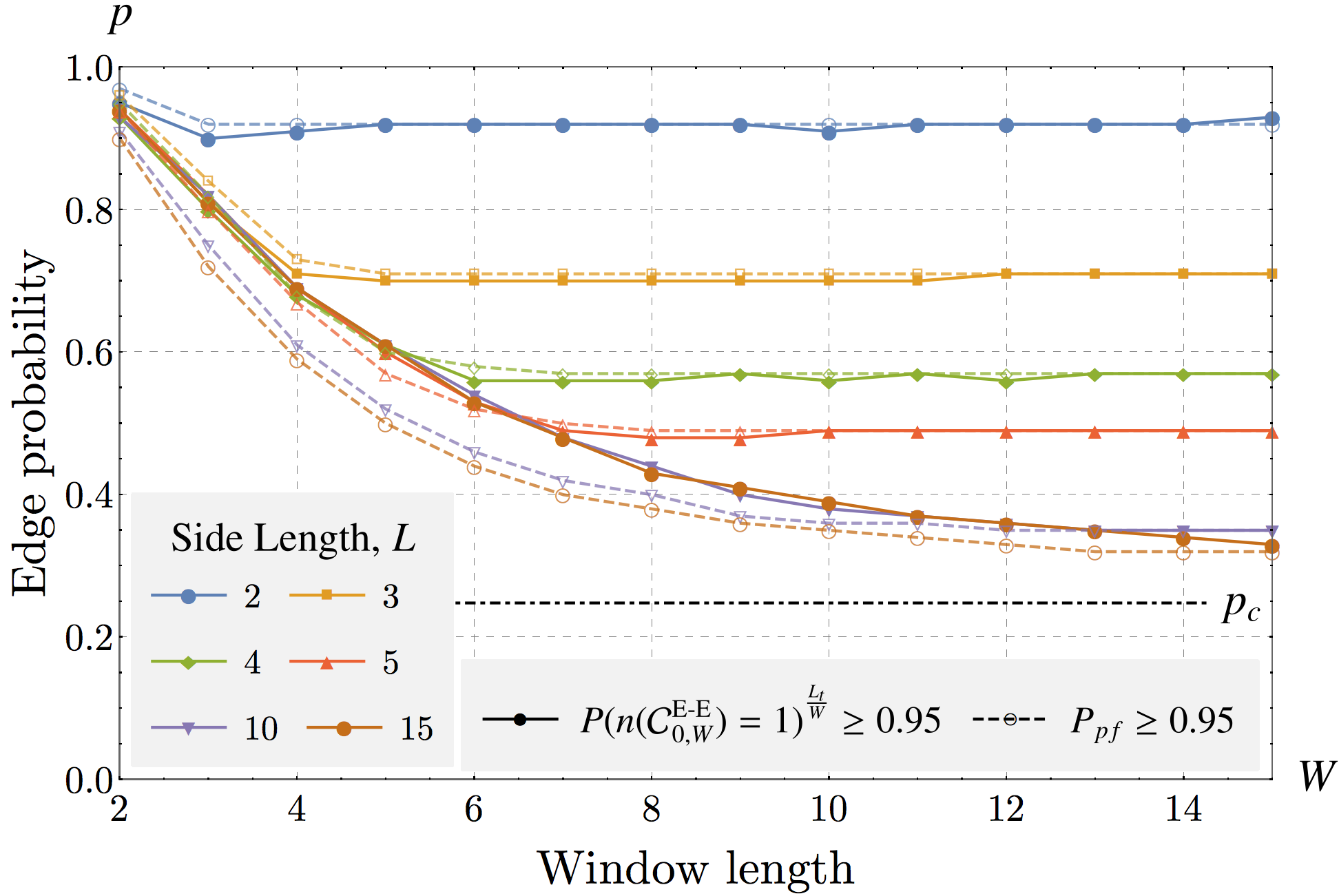}
	\caption{Contours for $P(n(\mathcal{C}^\emphrm{E-E}_{0,W})=1)^\frac{L_t}{W} \geq 0.95$ with $L_t = 1000$ (solid lines). Depicted in dashed lines are the contours depicted previously for successful LLP ($P_\emph{pf} \geq 0.95$) in Fig.\ \ref{fig:cubic_pathfinding_contour}.}
	\label{fig:cubic_pathf_and_perc_contours}
	\vspace{-5pt}
\end{figure}

In Sec.\ \ref{sec:pathf_sim} we showed that standard block percolation rates $P_t$ could be used to approximate LLP and long-range percolation in the region of $W \geq W_\emphrm{max}$ by use of a stacked block percolation model.
To extend this heuristic to our conjectured condition for successful LLP, we consider the approximation 
\begin{align}
P_\emphrm{pf} \approx P(n(\mathcal{C}^\emphrm{E-E}_{0,W})=1)^\frac{L_t}{W}.
\end{align}
This approximation allows us to consider the contours of $P(n(\mathcal{C}^\emphrm{E-E}_{0,W})=1)^\frac{L_t}{W} \geq 0.95$ that can be compared with those previously found for LLP.
To assess the validity of such an approximation both contours are depicted in Fig. \ref{fig:cubic_pathf_and_perc_contours}.
We find that this heuristic shows good agreement for small side lengths $L=2,3,4,5$ across all $W$, but underestimates pathfinding performance for larger side lengths $L=10, 15$ within the same region (although this is recovered for $W \geq W_\emphrm{max}$).
In an attempt to understand the discrepancy observed for large $L$ and small $W$, we consider two candidate explanations.

Firstly, it may be the case for thin but wide (low $W$, large $L$) active blocks that approximating LLP as stacked instances of block percolation is simply not a good model.
This may be due the the main simplification of this model: the lack of requirement for overlapping paths between blocks, an inaccuracy that becomes more pertinent as $W$ is deceased below $W_\emphrm{max}$.
However, one would expect such a simplification to over-estimate LLP performance, as observed for low $L$ and $W$.
Equally, it is also possible that our chosen feature function $F(\mathcal{B}_{t,t+W})$ based on $n(\mathcal{C}^\emphrm{E-E}_{t,t+W})$ cannot be equally applied across all LLP regimes, or that it is not valid to extend such statistics using a simple stacked-block approximation.

Secondly, this discrepancy could also be a result of perturbations from expected percolation statistics because of boundary effects.
This is potentially due to a subtlety in the definition of our LLP feature function and how it is approximated from instances of block percolation.
Within our LLP model, we strictly consider $\mathcal{C}^\emphrm{E-E}_{t,t+W} = \mathcal{B}_{t,t+W} \cap \mathcal{C}_{0,L_t}^\emphrm{E-E}$, which excludes end-to-end components in $\mathcal{B}_{t,t+W}$ that aren't part of the full (and unique) spanning cluster.
However, to numerically simulate $n(\mathcal{C}^\emphrm{E-E}_{t,t+W})$, instances of $\threed{W}{L}{L}$ cubic lattice $\mathcal{B}_{0,W}$ were generated and the number of connected end-to-end components $n(\mathcal{C}^\emphrm{E-E}_{0,W})$ found. 
The statistics we find are hence only strictly equivalent to LLP for $L_t = W$; an exact simulation would require a simulation that: generates the full $\threed{L_t}{L}{L}$ lattice, extracts the unique end-to-end component, and then finds $n(\mathcal{C}^\emphrm{E-E}_{t,t+W})$ for all times $0 \geq t \geq L_t - W$ (or a random selection thereof).
When simulated, the likelihood of finding $n(\mathcal{C}^\emphrm{E-E}_{0,W}) > 1$ is therefore increased for blocks with both small $W < W_\emphrm{max}$ and large $L=10, 15$  as additional structure is considered that would have otherwise been ignored in LLP (representing end-to-end components in $\mathcal{B}_{t,t+W}$ that are disjoint from $\mathcal{C}_{0,L_t}^\emphrm{E-E}$).
If such structures are present, this would work to explain the underestimation of LLP performance, as well as highlighting the limitations simulating LLP with block percolation for $W < W_\emphrm{max}$.
This discrepancy emphasises the importance of considering unexpected boundary effects when modelling percolation statistics, especially for small lattices with high surface to volume ratio.

\end{document}